\documentclass[acus]{jacow}

\usepackage{graphicx}
\usepackage{booktabs}
\usepackage{url}
\usepackage{cite}
\usepackage[caption=false]{subfig}

\begin{document}
\title{PROGRESS ON BEAM-PLASMA EFFECT SIMULATIONS IN MUON IONIZATION COOLING LATTICES\thanks{Work supported by the U.S. Department of Energy.}}

\author{J. Ellison, P. Snopok, Illinois Institute of Technology, Chicago, IL, USA}

\maketitle

\begin{abstract}
New computational tools are essential for accurate modeling and simulation of the next generation of muon-based accelerators. One of the crucial physics processes specific to muon accelerators that has not yet been simulated in detail is beam-induced plasma effect in liquid, solid, and gaseous absorbers. We report here on the progress of developing the required simulation tools and applying them to study the properties of plasma and its effects on the beam in muon ionization cooling channels.
\end{abstract}

\section{Introduction}
Though muon accelerator simulation codes have been steadily improving over the years, there is still much room for improvement. Many single-particle processes and collective effects in vacuum and matter, such as space charge, beam-beam effects, plasma effects from ionized electrons and ions have not been implemented in any single current code. These effects have to be either deemed negligible or taken into account to ensure the proper accuracy of simulations.

\begin{figure}[htb]
\centering
\includegraphics[width=0.48\textwidth]{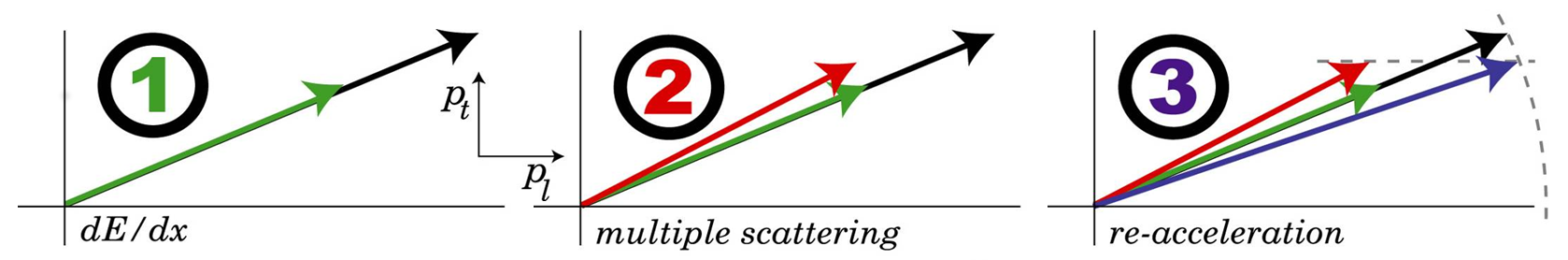}
\caption{Principle of ionization cooling: 1) The overall momentum is reduced through ionization where $\left\langle\frac{dE}{dx}\right\rangle$ is the mean energy loss of the muons. 2) Transverse momentum increases through multiple scattering. 3) Through re-acceleration, longitudinal momentum is regained.}
\label{simscreenshot}
\end{figure}

Ionization cooling (principle illustrated in Fig.~\ref{simscreenshot}) is a method by which the emittance of a muon beam can be reduced. A beam is sent through a material, losing momentum through multiple scattering and ionization processes, and reducing its emittance. By re-accelerating the beam through RF cavities, the longitudinal momentum is restored, and any lost energy is regained so that the process can be repeated.

The evolution of the normalized transverse emittance can be described by the following equation:
\[
\frac{d\epsilon_n}{dz}\approx-\frac{1}{\beta^2}\left\langle\frac{dE_\mu}{dz}\right\rangle\frac{\epsilon_n}{E_\mu}+\frac{1}{\beta^3}\frac{\beta_\perp E_s^2}{2E_\mu mc^2X_0},
\]
where $\epsilon_n$ is the normalized emittance, $z$ is the path length, $E_\mu$ is the muon beam energy, $\beta=v/c$, $X_0$ is the radiation length of the absorber material, $\beta_\perp$ is the betatron function, and $E_s$ is the characteristic scattering energy~\cite{neuffer}. Here, two competing effects can be seen: the first term is the cooling (reduction of phase space beam size) component from ionization energy loss and the second term is the heating (increase of phase space beam size) term from multiple scattering. For minimizing heating, a small betatron function from a strong magnetic field and a large radiation length are needed. To maximize cooling, a large stopping power is needed, $\left\langle\frac{dE_\mu}{dz}\right\rangle$. Hydrogen gives the best balance between a large radiation length and a large stopping power.

\begin{figure}[htb]
\centering
\includegraphics[width=0.48\textwidth]{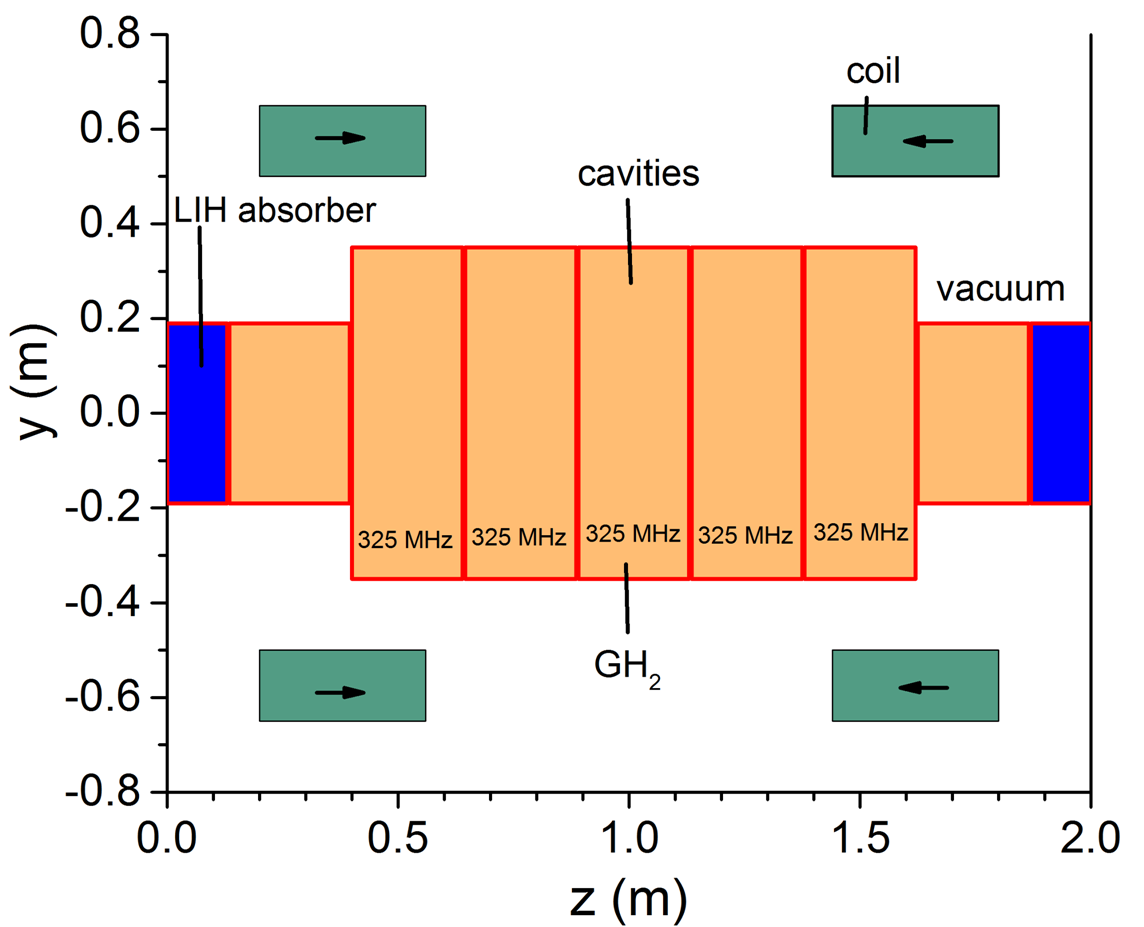}
\caption{Example of a matter-dominated hybrid cooling channel with gas-filled RF cavities, nearly all the cell length has material: either medium-pressure gaseous hydrogen or LiH absorber. (Image courtesy of D. Stratakis)}
\label{sample_cell}
\end{figure}

Muons will ionize material as they travel through absorbers. This will generate a plasma, and it is the interaction of the muon beam with the generated plasma that is studied here. Beam-plasma interaction is not taken into account currently in a majority of muon accelerator simulation codes. This interaction is especially important when simulating ionization cooling in the hybrid cooling channels with medium-to-high pressure gas-filled RF cavities (Fig.~\ref{sample_cell}). 

The plasma effects have been studied by plasma physicists, but have not been studied extensively from a beam physics point of view. The plasma has been shown not to disrupt the beam or make it blow up dramatically~\cite{Ahmed}, however, for ionization cooling purposes beam-plasma effects may have a large impact on the cooling rates for both charges of muons. Essentially, the head of a bunch sees a material with different properties than the tail of the bunch and whole bunches may see materials with different properties than the previous bunches. Ionization rates vary from material to material so the effects may be more prominent in some materials than others.

\section{Simulations}
After several simulation packages were considered, the one found to best suit our needs was WARP~\cite{WARP}. WARP is an actively developed particle-in-cell (PIC) simulation code designed to simulate particle beams with high space-charge intensity.


Several ionization models were considered to generate the plasma including multiple ways to introduce the plasma manually, but ultimately the ionization module contained within WARP was used. Starting with an ionization cross section,
\[\sigma=\left\langle\frac{dE}{dx}\right\rangle\frac{1}{W_i}\frac{\rho}{\rho_n}.\]
WARP will generate the plasma on its own. Here, $\left\langle\frac{dE}{dx}\right\rangle$ is the mean rate of energy loss by the muons, $W_i$ is the average energy to produce an ion pair, and $\rho$ and $\rho_n$ are the mass and atomic densities of the medium, respectively.



In the proof-of-concept simulations with a dense muon beam, it was seen that beam-plasma effects can significantly alter the results. The bunch shape varied drastically when comparing the simulation results with and without plasma effects~\cite{Ellison}. WARP proved to calculate the desired effects fairly efficiently, with a factor of six slow down when plasma effects were included.





\section{Progress on simulations}
Beam-plasma effects have been shown to potentially have a significant impact on the shape of a muon bunch. This impact needs to be quantified, and the effect on cooling rates needs to be studied. To do this, a section of a realistic cooling channel has to be simulated.

In the previous simulations, scattering and straggling have not been taken into account due to the lack of these features in WARP. Subsequently, a WARP-ICOOL wrapper has been used~\cite{Grote}, incorporating into WARP the scattering and straggling processes from ICOOL~\cite{ICOOL}. At the end of each step inside the material, WARP calls the relevant ICOOL processes and applies them to particles in the simulation.  

\begin{figure}[t!]
	
	\subfloat[][6D emittance]{%
	    \includegraphics[clip,width=\columnwidth]{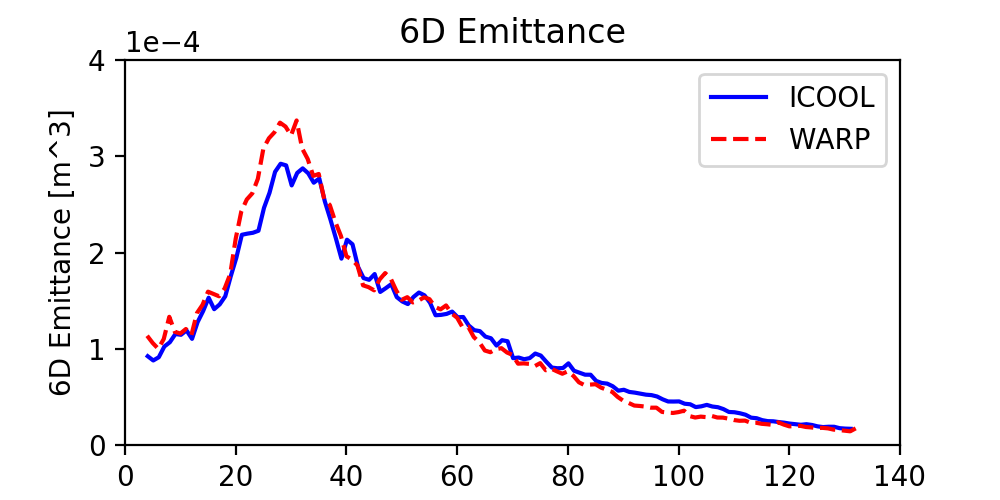}%
	}
	
	\subfloat[][longitudinal emittance]{%
		\includegraphics[clip,width=\columnwidth]{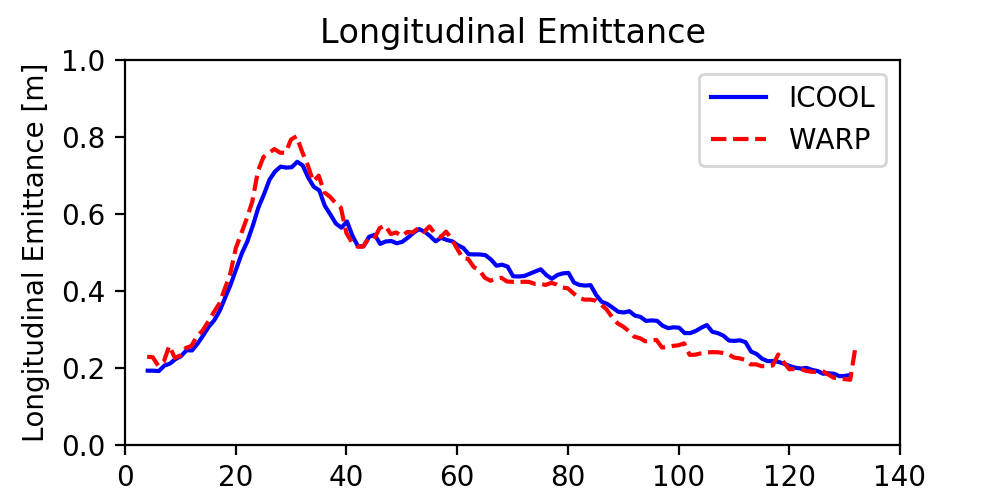}%
	}
	
	\subfloat[][transverse emittance]{%
		\includegraphics[clip,width=\columnwidth]{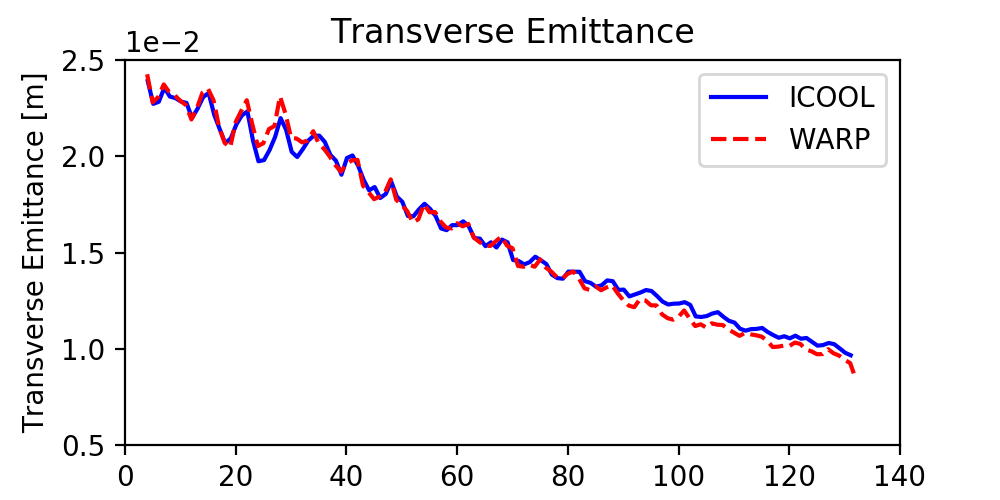}%
	}
	
	
	\caption{Comparison of the six-dimensional (a), longitudinal (b), and transverse (c) emittances in ICOOL (blue) and WARP (red) simulations with stochastic effects through the first stage of the rectilinear cooling channel before bunch merging.} 
	\label{Stochastic_plots}
\end{figure}

A complete cooling cell based on the first stage of the current version of the rectilinear cooling channel~\cite{RCC} has been modeled with the layout similar to two of the cells in Figure~\ref{sample_cell}, one after another, with magnetic field polarity reversed in the second half. This cell consists of four tilted solenoidal coils producing a maximum on-axis magnetic field of 2.36~T, six 325~MHz RF cavities with a maximum gradient of 22~MV/m and accelerating phase of 14\textdegree, and liquid hydrogen wedge absorber with a 35~\textdegree dispersion angle between solenoids. This stage of the cooling channel has been modeled in ICOOL, WARP, and G4beamline~\cite{G4beamline} without stochastic effects, using a representative beam sample, and all three simulations were in good agreement with one another. Stage 1 was then simulated in both ICOOL and WARP, both with stochastic effects. Results were compared using Ecalc9~\cite{ecalc9} and are summarized in Figure~\ref{Stochastic_plots}. Any differences can be attributed to these effects, verifying the validity of WARP simulations with this lattice. 



\begin{figure}[h]
	
	\subfloat[][6D emittance]{%
	    \includegraphics[clip,width=\columnwidth]{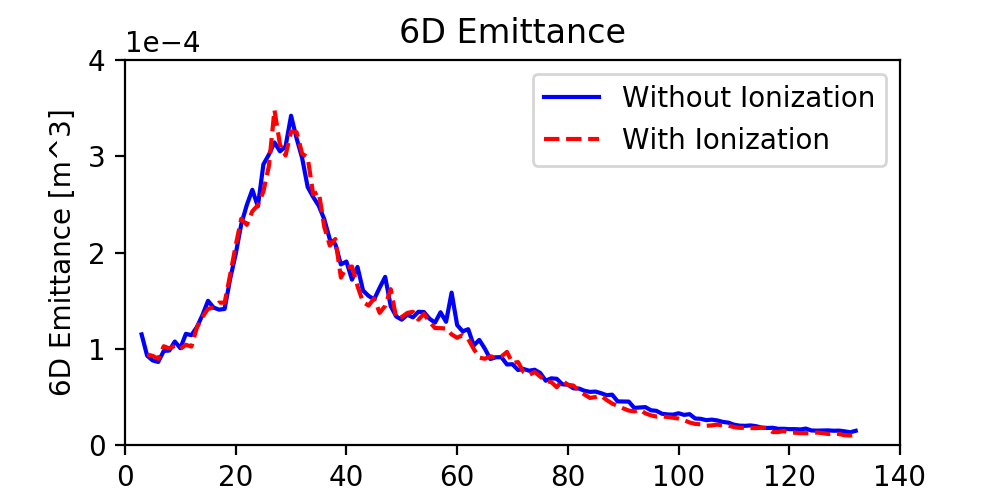}%
	}
	
	\subfloat[][longitudinal emittance]{%
		\includegraphics[clip,width=\columnwidth]{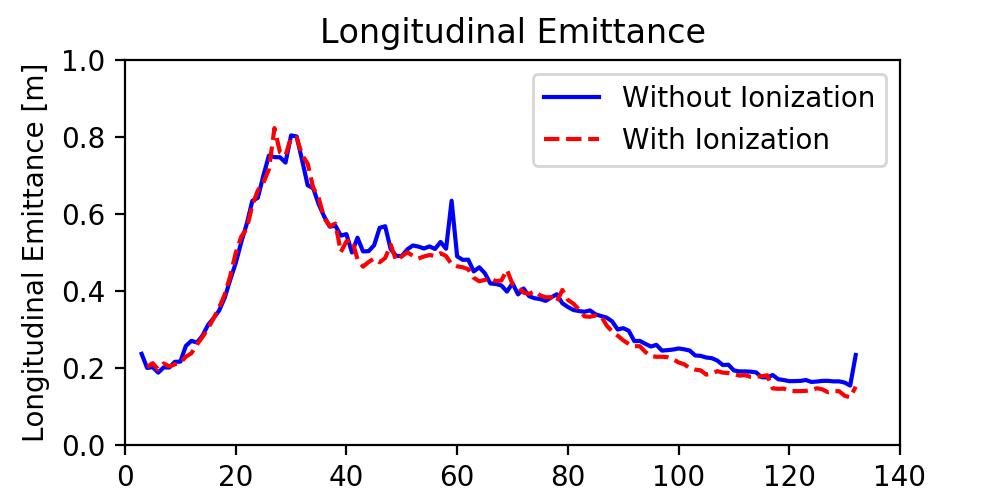}%
	}
	
	\subfloat[][transverse emittance]{%
		\includegraphics[clip,width=\columnwidth]{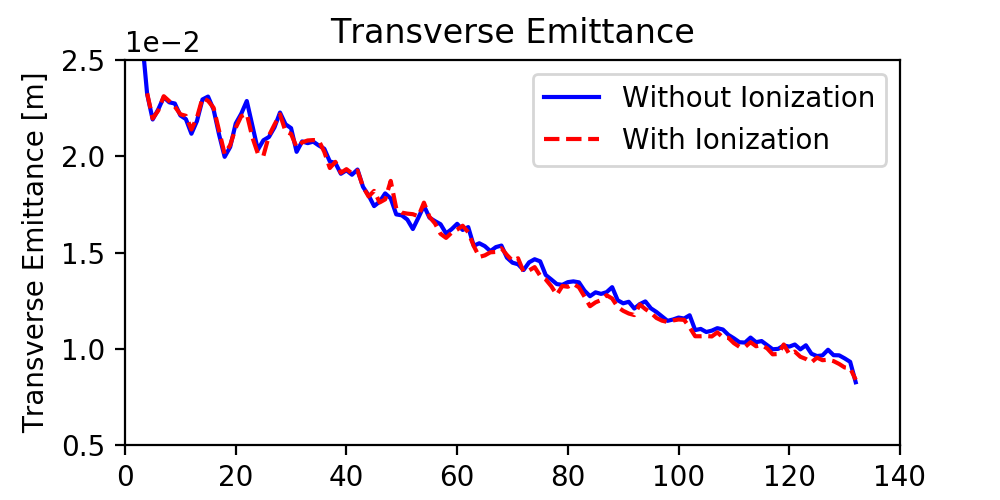}%
	}
	
	
	\caption{Comparison of the six-dimensional (a), longitudinal (b), and transverse (c) emittances in WARP simulations both without (blue) and with (red) ionization effects through the first stage of the rectilinear cooling channel before recombination.} 
	\label{ionization_plots}
\end{figure}

WARP simulations were then done through the same stage of the cooling channel, with stochastic effects, both with and without ionization. These are summarized in Fig.~\ref{ionization_plots}. The variations between these two can be contributed to both stochastic effects and beam-plasma interaction. In this early stage of the cooling process, the beam-plasma effects appear to be negligible.

\section{Current Challenges}

The results shown here are from a model based on the first stage of the proposed rectilinear cooling channel. The muon beam and electromagnetic fields will be much stronger after recombination and in later stages. It is expected the beam-induced plasma effects will be more substantial, and quantifiable.   

Several effects are still not taken into account here, including beam-plasma effects on subsequent bunches and plasma-ion recombination. These are currently being investigated, along with expanding the limits of the simulation to include multiple stages.

\section{ACKNOWLEDGMENT}
Authors would like to thank Alvin Tollestrup, Ben Freemire, and Katsuya Yonehara for many fruitful discussions, along with David Grote and Jean-Luc Vay for assistance with the WARP simulation package they (and others) developed.

This research used resources of the National Energy Research Scientific Computing Center, a DOE Office of Science User Facility supported by the Office of Science of the U.S. Department of Energy under Contract No. DE-AC02-05CH11231.

\end{document}